\documentclass[12pt]{iopart}
\usepackage{graphicx}

\expandafter\let\csname equation*\endcsname\relax
\expandafter\let\csname endequation*\endcsname\relax
\usepackage{amsmath}

\usepackage[normalem]{ulem}

\usepackage{color}

\newcommand{\kt}{k_{\text{B}}T}
\usepackage{hyperref}
\providecommand{\keywords}[1]{\textbf{Keywords:} #1}

\begin{document}
\title[Determination of folding free energies in single-molecule experiments]{Efficient methods for determining folding free energies in single-molecule pulling experiments}

\author{A Severino$^1$, A M Monge$^1$, P Rissone$^1$ and F Ritort$^{1,2}$}

\address{$^1$ Dept. F\'isica de la Mat\`eria Condensada, Universitat de Barcelona, C/ Mart\'i i Franqu\`es 1, 08028 Barcelona, Spain}
\address{$^2$ CIBER-BBN de Bioingenier\'ia, Biomateriales y Nanomedicina, Instituto de Salud Carlos III, 28029 Madrid, Spain}

\ead{fritort@gmail.com}

\begin{abstract}
The remarkable accuracy and versatility of single-molecule techniques make possible new measurements that are not feasible in bulk assays. Among these, the precise estimation of folding free energies using fluctuation theorems in nonequilibrium pulling experiments has become a benchmark in modern biophysics. 
In practice, the use of fluctuation relations to determine free energies requires a thorough evaluation of the usually large energetic contributions caused by the elastic deformation of the different elements of the experimental setup (such as the optical trap, the molecular linkers and the stretched-unfolded polymer). We review and describe how to optimally estimate such elastic energy contributions to extract folding free energies, using DNA and RNA hairpins as model systems pulled by laser optical tweezers.  The methodology is generally applicable to other force-spectroscopy techniques and molecular systems.  
\end{abstract}

\maketitle

\section{Introduction}
Predicting free-energy differences is a central problem in molecular biophysics. Protein folding \cite{aabert2002molecular}, DNA hybridization \cite{felsenfeld1967physical}, ligand binding, CRISPR--Cas9 RNA editing  \cite{makarova2011evolution,sander2014crispr}, are molecular reactions whose fate is determined by the free-energy difference between reactants and products. Finding methods to extract free-energy, enthalpy and entropy differences is an essential task in biochemistry, where most of these quantities are measured by employing bulk techniques such as calorimetry, UV absorbance, fluorescence, surface plasmon resonance, among others \cite{cantor1980biophysical}. Bulk methods yield results that are incoherent temporal averages over a large population of molecules that are in different states. The signal is masked by the dominant species and reactions, limiting the capability of detecting rare non-native states and reaction pathways. Moreover, bulk molecular transformations often exhibit strong hysteresis effects rendering equilibrium differences inaccessible. 

By monitoring molecules one at a time, techniques such as single-molecule fluorescence \cite{shashkova2017single}, single-molecule translocation across nanopores \cite{meller2001voltage} and single-molecule force spectroscopy \cite{neuman2008single} overcome the previous limitations and therefore have become key experimental tools in many laboratories worldwide. In particular, force-spectroscopy techniques using atomic-force microscopy, magnetic tweezers, acoustic-force spectroscopy and laser optical tweezers (LOT) have been extremely fruitful, revolutionizing biophysics over the last three decades\footnote{LOT invention revealed to be a breakthrough in laser physics and has been awarded with the Nobel Prize in Physics in 2018 \cite{ashkin20182018}.}.

The main advantage of force-measuring techniques (as compared to fluorescence and other non-invasive optical technologies) lies in the possibility to measure simultaneously force and displacement, giving direct access to mechanical work measurements in single-molecule pulling experiments. Similarly to bulk assays, pulling experiments are often carried out under irreversible conditions, in principle providing bounds (rather than direct estimates) of equilibrium free-energy differences. The development of the non-equilibrium thermodynamics of small systems (also known as stochastic thermodynamics) \cite{bustamante2005nonequilibrium,seifert2008stochastic,seifert2012stochastic,ciliberto2017experiments} during the past three decades has provided the theoretical concepts and methods needed to extract free-energy differences from repeated irreversible work measurements. Exact results such as the Jarzynski equality \cite{jarzynski1997on} and the Crooks fluctuation theorem \cite{crooks2000path} are now commonly employed to extract free-energy differences from single-molecule pulling experiments \cite{liphardt2002equilibrium,collin2005verification,ritort2008nonequilibrium,hummer2010free,gupta2011experimental,ross2018equilibrium}. A particularly useful application is the measurement of the folding free energy of nucleic acids and proteins ($\Delta G_0$) which is equal to the free energy difference between the folded structure and the unstructured random coil in the solvent. This quantity can be obtained from pulling experiments by measuring the free energy difference ($\Delta G$) between the folded and unfolded-stretched states of the considered experimental system taken at two force values, and by deriving from it the value of $\Delta G_0$.  However, a general problem in the manipulation of small systems using single-molecule techniques is that we cannot abstract away certain components or parts -- generally known as instrumental artifacts -- of the full experimental system. Hence, the quantity that can be obtained from pulling experiments using non-equilibrium thermodynamics is not $\Delta G_0$ directly. It is instead the free energy difference $\Delta G$ between the folded and unfolded-stretched states of the \textit{entire} considered experimental system taken at two force values. In order to retrieve the 'bare' molecular properties such as the value of $\Delta G_0$ in a single molecule, we therefore need to retrench from $\Delta G$ some contributions stemming from the experimental set-up (e.g. optical trap in LOT or cantilever in AFM and the linkers used to manipulate the molecule under study). These so-called stretching corrections play a crucial role because their contribution to the total free energy difference $\Delta G$ are much larger than the free energy one wants to extract $\Delta G_0$, making the accurate estimation of the latter a difficult task. Although there are several studies on the influence of the instrumental artifacts on the folding kinetics in single-molecule experiments \cite{dudko2008theory,maitra2010model,maitra2011influence,berkovich2012rate,chang2013effect,hinczewski2013mechanical,cossio2015artifacts,neupane2016quantifying}, their influence regarding the determination of the folding free energies at zero force has, to the best of our knowledge, never been addressed in detail.

In this work we will rigorously examine these experimental contributions in LOT showing how to efficiently and reliably estimate the free energies of formation of DNA and RNA hairpins in unzipping assays. The same methodology is applicable to proteins and ligand binding interactions using LOT or other force measuring techniques as well (AFM, magnetic tweezers and so on). The development of novel and refined statistical analysis methods to extract differences in thermodynamic potentials (free energy, enthalpy, entropy, chemical potential, ...) will become crucial with the recent boost of high-throughput single-molecule techniques (magnetic tweezers, acoustic force spectroscopy) that will require fast and efficient algorithms. 

The content of this paper is organized as follows. In sections \ref{sec:model} and \ref{sec:stretching} we describe the typical experimental setup of LOTs and then define and discuss the different contributions to the total free energy. The two following sections (\ref{sec:effectivestiffness-examples} and \ref{sec:beyond-effective-stiffness}) feature how to estimate these contributions when analyzing DNA and RNA molecules. Section \ref{sec:effectivestiffness-examples} first covers the situations in which it is possible to introduce the so-called effective stiffness approximation, which considerably simplifies the computation of the large stretching terms. When this approximation fails, a more elaborate approach requiring a careful evaluation of the elastic response of the linkers and of the force probe is needed, and this is the focus of section \ref{sec:beyond-effective-stiffness}. Finally, in section \ref{sec:conclusions} we present the conclusions.

\section{Model of the experimental setup}\label{sec:model}

\begin{figure}[t]
\centering
\includegraphics[width=\textwidth]{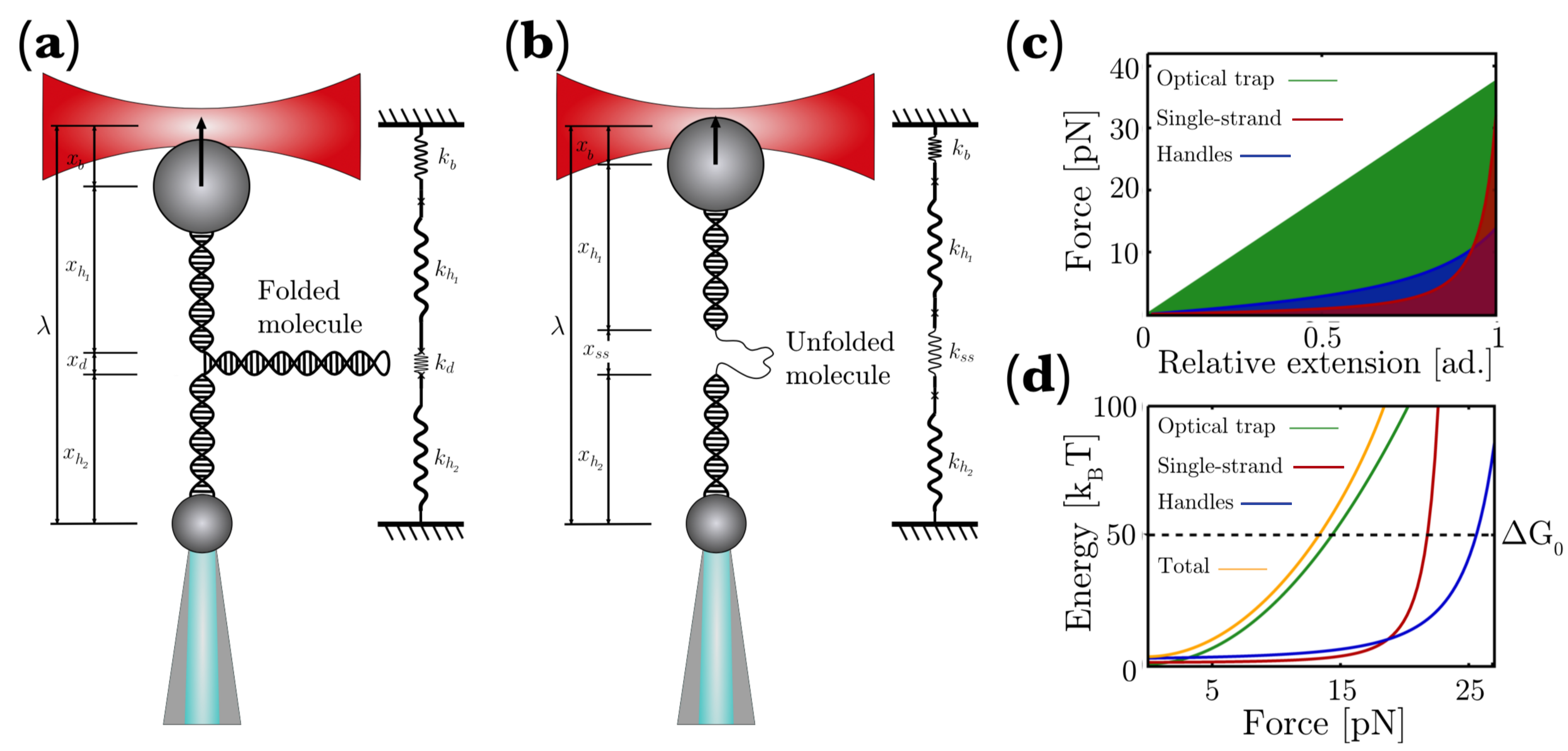}
\caption{\label{fig:CartoonExt}\textbf{(a,b).} {\bf Laser optical tweezers (LOT) experimental setup.} The molecule is tethered between two polystyrene beads using two dsDNA (or dsRNA or even dsDNA/DNA hybrids) handles. Arrow towards the center of the optical trap indicates the direction of the force. $\lambda = x_{\text{b}} + x_{\text{h}} + x_{\text{m}}$ (with $x_{\text{h}} = x_{\text{h}_1} + x_{\text{h}_2}$) is the relative distance between the center of the optical trap and the tip of the micropipette. $x_{\text{m}}$ equals $x_{\text{d}}$ when the molecule is folded (a) or $x_{\text{ss}}$ when the molecule is unfolded (b). \textbf{(c).} Sketch of the force versus relative extension (extension divided by contour length) for each elastic element  showing their respective energy contributions (shaded areas).  \textbf{(d).} Elastic energy contribution of each element vs force and comparison with the typical energy of formation (dashed line, $\Delta G_0$) for a 20bp DNA or RNA hairpin.}
\end{figure}

We consider the case of a nucleic acid (DNA or RNA) hairpin pulled by LOT. In LOT, the total distance $\lambda$ between the tip of the micropipette and the center of the optical trap is the control parameter of the experiments. As shown in figure \ref{fig:CartoonExt}(a),(b) the distance $\lambda$ can be decomposed as:

\begin{equation}\label{eq:ext_tot}
\lambda (f) = \begin{cases} 
    x_{\text{b}}(f) + x_{\text{h}}(f) + x_{\text{d}}(f)+{\rm const} \, \text{ (folded state)} \, , \\
    x_{\text{b}}(f) + x_{\text{h}}(f) + x_{\text{ss}}(f) +{\rm const}\,  \text{ (unfolded state)} \, ,
    \end{cases}
\end{equation}
depending on whether the molecule is folded or unfolded. Here $x_{\text{b}}(f)$ is the displacement of the bead from the center of the optical trap, $x_{\text{h}}(f) = x_{\text{h}_1}(f) + x_{\text{h}_2}(f)$ accounts for the sum of the elongations of the two double-stranded handles,  $x_{\text{ss}}(f)$ is the end-to-end extension of the single-stranded unfolded molecule, and $x_{\text{d}}(f)$ is the average extension of the folded hairpin. This last term is defined as the distance between the attachment points of the handles to the 5' and 3' ends of the hairpin and is usually called 'hairpin diameter' (whence the index $d$). All these extensions are evaluated against the $x$-(pulling)axis and at a given force $f$. The '${\rm const}$' stands for an arbitrary shift in the total distance $\lambda$ which does not affect the analysis.  

In general, a small perturbation $\delta \lambda$ generates a small change in the applied force $\delta f$. The extent of this variation is the \emph{effective stiffness} of the system $k_{\text{eff}}= \delta f / \delta \lambda$ and it equals the slope of the experimental force-distance curve (FDC). 
Therefore, according to the above definition and to the prescription given in \eqref{eq:ext_tot}, the inverse effective stiffness of the folded (F) and unfolded (U) branches are respectively given by:

\begin{subequations}
\begin{align}
   \frac{1}{k_{\text{eff}}^{\text{F}}(f)} =  \frac{1}{k_{\text{b}}(f)} +\frac{1}{k_{\text{h}}(f)} + \frac{1}{k_{\text{d}}(f)} \, , \label{eq:keffBranches} \\
    \frac{1}{k_{\text{eff}}^{\text{U}}(f)} =  \frac{1}{k_{\text{b}}(f)} +\frac{1}{k_{\text{h}}(f)} + \frac{1}{k_{\text{ss}}(f)} \, .
\end{align}
\end{subequations}

where $k_{\text{b}}(f)$ corresponds to the stiffness of the bead in the optical trap, $k_{\text{h}}(f)$ is the sum of the two handles stiffness and $k_{\text{d}}(f)$, $k_{\text{ss}}(f)$ stand for the molecular stiffness of the folded and unfolded molecule, respectively.

In particular, $k_{\text{d}}(f)$ is modelled as the stiffness to orient a dipole  of diameter $d$ (typically $d = 2$ nm for DNA and RNA hairpins \cite{woodside2006direct}) along the force axis \cite{forns2011improving}. Recalling that in general $k^{-1}=\delta x/\delta f$, the dipole stiffness can be derived from the well-known relation between a dipole average extension (which is here equal to the average extension of the folded hairpin) and the force $f$ to which it is subjected:

\begin{equation}\label{eq:fjc}
x_{\text{d}}(f) = d \left[ \coth \left( \frac{fd}{\kt}\right) - \frac{\kt}{fd} \right] \, 
\end{equation}

where $T$ is the temperature of the heat bath around the dipole and $k_{\mathrm{B}}$ is Boltzmann constant.

An analytic expression for $k_{\text{ss}}$ and $k_{\text{h}}$ can be obtained by describing the elastic response of nucleic acids in their single-stranded and double-stranded form with the Worm-Like Chain (WLC) polymer model and its interpolation formula \cite{marko1995stretching},

\begin{equation}\label{eq:wlc}
f(x) = \frac{\kt}{4P} \left[ \left(1-\frac{x}{L_{c}} \right)^{-2} - 1 + 4 \frac{x}{L_{c}} \right] \, 
\end{equation}

where $x$ is the average extension of the molecule ($x=x_{\text{ss}}$ for the unfolded hairpin, $x=x_{\text{h}}$ for the double-stranded handles) and $P$ is the persistence length, i.e. the typical distance along the polymer backbone over which there is an appreciable bending due to thermal fluctuations. $L_{c}$ is the contour length, i.e. the end-to-end distance of the fully straightened polymer, which can also be written as $L_c = n d_{\text{b}}$ with $n$ being the total number of monomers in the polymer and $d_{\text{b}}$ the length per monomer. In general, inverting \eqref{eq:wlc} to get $x(f)$ is not an easy task (the full computation is reported in the \ref{app:inversionWLC}) and the solution depends on the system parameters. 

Finally, the stiffness of the polymer can be obtained by differentiation of \eqref{eq:wlc}:

\begin{equation}\label{eq:WLC_stiffness}
k(x) \equiv \frac{\partial f(x)}{\partial x} = \frac{k_BT}{2L_cP} \left[ \left(1-\frac{x}{L_{c}} \right)^{-3} + 2\right] \, .
\end{equation}

Given \eqref{eq:wlc}, it is also possible to further take into account the elastic deformation of the stretched polymer by performing the substitution $L_c\to L_c(1+f/Y)$, with $Y$ the Young modulus of the stretchable polymer \cite{wang1997stretching,petrosyan2017improved}, i.e. the resistance to deformation of the system to an applied uniaxial stress. In this case the contour length becomes force-dependent and the corresponding model is called the \emph{extensible} WLC. By contrast equation (4) where $L_c$ is constant is known as the \emph{inextensible} WLC. The latter has been shown to describe the elastic properties of single-stranded nucleic acids (ssDNA and ssRNA) with good accuracy \cite{camunas2016elastic} while the former has for long been the standard to model the elastic properties of double-stranded nucleic acids in the entropic regime.

The persistence length $P$ is a measure of the mechanical stiffness of the polymer being strongly sensitive to environmental conditions (e.g. ionic strength, temperature, solvation, etc..). Polymers with $P \gg L_{c}$ effectively behave as rigid rods, whereas if $P \leq L_{c}$ polymers are bent at the scale of the contour length by thermal forces. It is important to mention that $P$ does not only depend on the ionic concentration and temperature \cite{de2015temperature} (as predicted by polyelectrolyte theories) but also on experimental parameters such as contour length \cite{camunas2016elastic}. For example, at 1 M NaCl, recent single-molecule studies have shown that, for short (a few tens bases) ssDNA molecules, $P = 1.35$ nm \cite{alemany2014determination} whereas for long ssDNAs $\sim$ 13 kb $P = 0.76$ nm \cite{bosco2013elastic}. On the other hand, for short ssRNA molecules $P = 0.75$ nm \cite{bizarro2012non} and for long $\sim$ 1 kb ssRNAs $P = 0.83$ nm \cite{camunas2016elastic}.  These values are significantly lower than for double-stranded nucleic acids (dsDNA and dsRNA) where $P = 50$ nm for dsDNA \cite{bustamante1994entropic} and $P \simeq 60$ nm for dsRNA molecules \cite{abels2005single}.

\section{Stretching contributions and free-energy recovery}\label{sec:stretching}

Let us suppose that initially at $t = 0$ we have a molecule in thermal equilibrium at the folded (or native, $N$) state at a given value $\lambda_{0}$ of the control parameter. Then, we perturb the system by applying a predetermined time-dependent forward (F) protocol, $\lambda_{F}(t)$, that starting at $\lambda_{0}$ at $t=0$ ends at an arbitrary $\lambda_{1}$ at a time $t_1$. The mechanical work $W$ done along this process equals to:

\begin{equation}
\label{eq:work}
 W =  \int_{\lambda_0}^{\lambda_1} \, f d \lambda \, .
\end{equation}

The Crooks Fluctuation Theorem (CFT) \cite{crooks2000path} relates the mechanical work done on a system in a set of arbitrary irreversible measurements with the equilibrium free-energy difference of this system between $\lambda_0$ and $\lambda_1$, $\Delta G = G(\lambda_{1}) - G(\lambda_{0})$. It reads:

\begin{equation}
\label{eq:CFT}
\frac{P_{F}(W)}{P_{R}(-W)} = \exp{\left( \frac{W - \Delta G}{\kt} \right)} \, ,
\end{equation}

where $P_{F}(W)$ is the probability distribution of the work done in the F process and $P_{R}(-W)$ is the probability distribution of the work measured in the time-reversed (R) process (i.e. starting in thermal equilibrium in $\lambda_{1}$ and performing the time-mirrored protocol so that: $\lambda_{R}(t) = \lambda_{F}(t_1-t)$). The derivation of the CFT has become a milestone for single-molecule experimentalists, allowing the measurement of free-energy differences in conditions where traditional bulk experiments are unfeasible.  By pulling single molecules using LOT or magnetic tweezers it is possible to recover molecular free-energy differences from irreversible work measurements \cite{collin2005verification,monge2018experimental}. The CFT (Eq.\ref{eq:CFT}) implies the well-known Jarzynski equality \cite{jarzynski1997on}:

\begin{equation}
\label{eq:JE}
\left\langle \exp{\left( - \frac{W }{\kt} \right)} \right\rangle_{F} = \exp{\left( -\frac{ \Delta G}{\kt} \right)} \, ,
\end{equation}

Note that the average $\langle \cdots \rangle_{F}$ is evaluated over $P_{F}(W)$ (an analogous equality holds for the reverse process). It is important to bear in mind that the free energy $\Delta G$ obtained using the CFT 
\eqref{eq:CFT} (or the Jarzynski equality \eqref{eq:JE}) contains several contributions due to the stretching of the different parts forming the experimental setup. These are the molecule under study, the molecular handles and the optically trapped bead (figure \ref{fig:CartoonExt}(a),(b)):

\begin{equation}\label{eq:gtot}
\Delta G = \Delta G_0 + \Delta W_{\text{m}} + \Delta W_{\text{b}} + \Delta W_{\text{h}}.
\end{equation}

$\Delta G_0$ is the free energy of formation of the molecule at zero force, which is equal to the free energy difference between the folded and unfolded hairpin conformations in solution (i.e. without optical trap and handles and without any applied force). The quantities $\Delta W_{i}$ ($i = \text{m,\,b,\,h}$) are the reversible work differences between the state of the $i^{th}$ setup element (optical trap, handles or molecule) at $\lambda_{0}$ (where the hairpin is folded and subjected to a minimum force $f_{\text{min}}$) and $\lambda_{1}$ (where the hairpin is unfolded and subjected to a maximum force $f_{\text{max}}$). Mathematical definitions of these quantities for the LOT setup are given in the subsections below.\\

As depicted in figure \ref{fig:CartoonExt}(c,d), for typical unfolding forces in DNA and RNA hairpins (15 - 25 pN), \eqref{eq:gtot} is dominated by the trap contribution, while the other terms have the same order of magnitude. Therefore, an accurate measurement of $\Delta G_0$ requires precise knowledge of all the different energetic contributions involved in the mechanical unfolding of the molecule. 

\subsection{Molecular stretching contribution} 

The molecular contribution $\Delta W_{\text{m}}$ in \eqref{eq:gtot} accounts for the reversible work needed to stretch the molecule under study and it can be written as:

\begin{equation}
\label{eq:wmol}
\Delta W_{\text{m}} = \int_{0}^{x_{\text{ss}}(f_{\text{max}})} f(x_{\text{ss}}) \, dx_{\text{ss}} -  \int_{0}^{x_{\text{d}}(f_{\text{min}})} f(x_{\text{d}}) \, dx_{\text{d}} \, ,
\end{equation}

where $f(x_{\text{ss}})$ and $f(x_{\text{d}})$ are the equilibrium force-extension curves of the unfolded and folded molecule, respectively (albeit different mathematical functions the same letter $f$ will be used to lighten the notation). The first term in the right-hand side of \eqref{eq:wmol} corresponds to the reversible work needed to stretch the unfolded molecule from its single-stranded random coil conformation at $f = 0$ up to $f_{\text{max}}$ and it can be computed from the WLC model, Eq. \eqref{eq:wlc}. The second term in the right-hand side of \eqref{eq:wmol} is the reversible work required to orientate the molecular diameter along the force axis. It can be written as:

\begin{equation}
\int_{0}^{x_{\text{d}}(f_{\text{min}})} f(x_{\text{d}}) dx_{\text{d}} = f_{\text{min}} \cdot x_{\text{d}}(f_{\text{min}}) - \int_{0}^{f_{\text{min}}} x_{\text{d}}(f) df \, .
\end{equation}

where $x_{\text{d}}(f)$ is given by \eqref{eq:fjc}.

\subsection{Bead and handles stretching contributions} 

The term $\Delta W_{\text{b}} + \Delta W_{\text{h}}$, which corresponds to the sum of the reversible work required to displace the bead from the center of the optical trap ($\Delta W_{\text{b}}$) plus the reversible work needed to stretch the handles ($\Delta W_{\text{h}}$), can be generally written as:
\begin{equation}
\begin{split}\label{eq:wbwh}
\Delta W_{\mathrm{b}} + \Delta W_{\mathrm{h}} &=  \int_{x_{\text{b}}(f_{\text{min}})}^{x_{\text{b}}(f_{\text{max}})} f(x_{\text{b}}) \, dx_{\text{b}} + \int_{x_{\text{h}}(f_{\text{min}})}^{x_{\text{h}}(f_{\text{max}})} f(x_{\text{h}}) \, dx_{\text{h}}\\
         & = \int_{f_{\text{min}}}^{f_{\text{max}}} f \left( \frac{\partial f}{\partial x_{\text{b}}} \right )^{-1} df + \int_{f_{\text{min}}}^{f_{\text{max}}} f \left( \frac{\partial f}{\partial x_{\text{h}}} \right )^{-1} df \\
         & = \int_{f_{\text{min}}}^{f_{\text{max}}} \frac{f}{k_{\text{b}}(f)} df + \int_{f_{\text{min}}}^{f_{\text{max}}} \frac{f}{k_{\text{h}}(f)} df \, .
\end{split}
\end{equation}

Note that each element in the setup is substantially different. In particular, the bead in the optical trap can be well approximated by a Hookean spring, whereas the elastic response of the handles and the single-stranded molecule (plus the diameter) is strongly nonlinear (see below). The contribution of these two terms in Eq.(\ref{eq:gtot}) is often large. In particular, the energy required to displace the bead from the center of the optical trap is considerably higher as compared to the other terms. A schematic depiction of this fact can be seen in figure \ref{fig:CartoonExt}(c), where the shaded areas below the curves represent the work $W$ obtained according to \eqref{eq:work} using realistic elastic parameters for DNA and RNA hairpins. 

\subsection{Effective stiffness approximation}\label{sec:effectivestiffness}
A further important simplifaction can be carried out when the FDC along the folded branch is approximately linear over the integration range of forces. Such a situation corresponds by definition to a scenario where the slope (or stiffness) is constant, i.e. $k_{\text{eff}}^{\text{F}} \neq k_{\text{eff}}^{\text{F}}(f)$. It allows one to readily perform the integration in eq. (\ref{eq:wbwh}) which is now reduced to the simple task  of integrating an affine function:
\begin{equation}
\label{eq:wbhandles2}
\Delta W_{\text{b}} + \Delta W_{\text{h}} = \int_{f_{\text{min}}}^{f_{\text{max}}} f \left( \frac{1}{k_\text{b}} + \frac{1}{k_\text{h}} \right) \, df \cong \int_{f_{\text{min}}}^{f_{\text{max}}} \frac{f}{k_{\text{eff}}^{\text{F}}} \, df = \frac{f_{\text{max}}^{2} - f_{\text{min}}^{2}}{2k_{\text{eff}}^{\text{F}}} \, ,
\end{equation}
where we used the fact that the stiffness of the dipole modelling the folded hairpin is considerably larger than the other terms in \eqref{eq:keffBranches}, so that $k^{\text{F}}_{\text{eff}} = (k^{-1}_{\text{h}} + k^{-1}_{\text{b}} + k^{-1}_{\text{d}})^{-1} \cong (k^{-1}_{\text{h}} + k^{-1}_{\text{b}})^{-1}$, and where the constant stiffness assumption is used in the last equality of the right hand side of (\ref{eq:wbhandles2}). Linearity of the FDC is a good approximation if the integration range is not too large (for example, when $f_{\text{max}}-f_{\text{min}} \approx 5 \; \text{pN}$, the case of the DNA and RNA hairpins considered in the next section.). Above all, linearity of the FDC certainly requires a linear optical trap of constant stiffness \cite{forns2011improving}. We will refer to this approximation as the \emph{effective stiffness method}. 

\section{The effective stiffness method}\label{sec:effectivestiffness-examples}

The effective stiffness approximation discussed in section \ref{sec:effectivestiffness} provides an easy method to treat the elastic contributions of the experimental setup. Here we provide two typical scenarios where \eqref{eq:wbhandles2} provides a reliable estimation of the free energy of formation $\Delta G_0$ of DNA and RNA hairpins. In section \ref{sec:cd4dna} the case of the  CD4 DNA hairpin with short handles is reported. Then in section \ref{sec:long_handles-CD4RNA} we discuss the case of the CD4 RNA  hairpin with long handles. 




\subsection{Short handles: the CD4 DNA hairpin.}
\label{sec:cd4dna}

\begin{figure}[t]
\centering
\includegraphics[width=\textwidth]{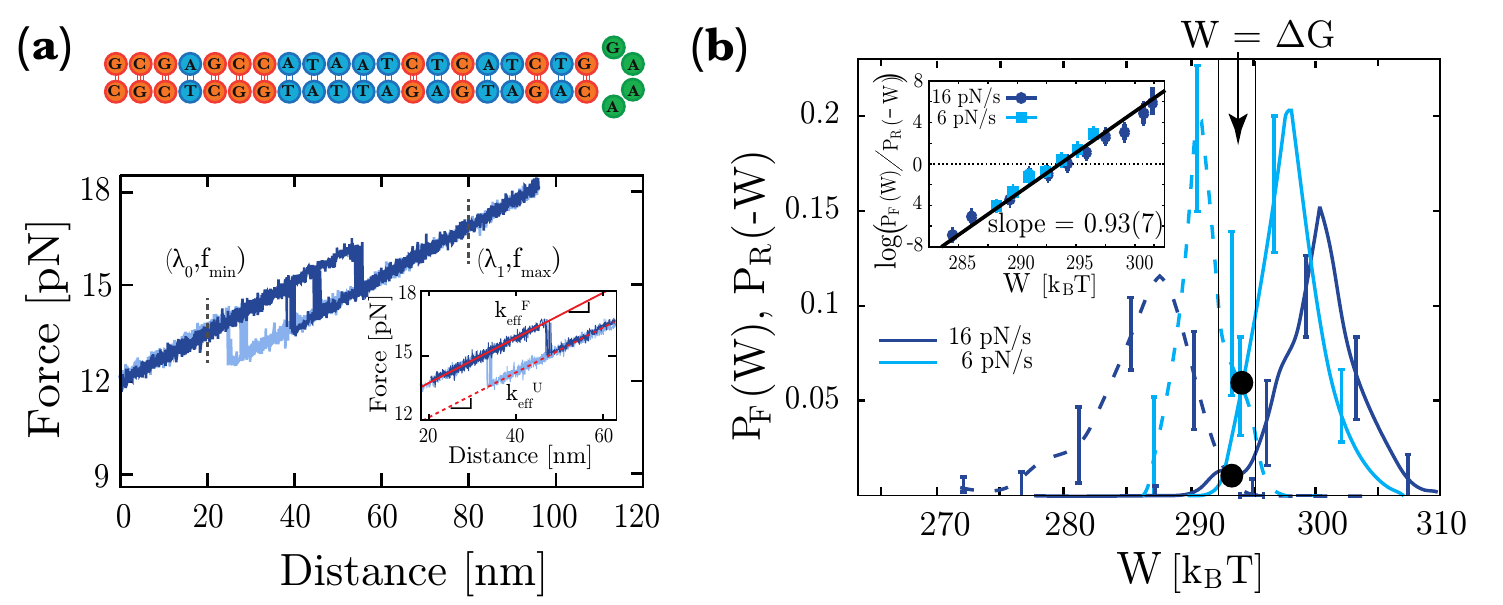}
\caption{\label{fig:CD4SH} \textbf{Free-energy recovery of CD4 DNA with short handles} \textbf{(a).} Sequence of CD4 DNA (top panel). FDCs and integration range for the work $W$ (bottom panel). Demonstration of the linearity of the FDCs in the integration range (inset) plus linear fits to the folded (solid line in the inset) and the unfolded branches (dashed line in the inset). \textbf{(b).} Forward (solid lines) and reverse (dashed lines) work distributions for two different pulling speeds calculated in the integration range indicated in (a) panel. Crossing points between work distributions are tagged as solid points. The CFT verification is shown as inset. Error bars have been obtained using the Bootstrap method.}
\end{figure}

The use of short dsDNA handles ($\sim$ 29 bp each) in single-molecule experiments has been shown to increase the precision of kinetic measurements due to their enhanced signal-to-noise ratio as compared to long handles \cite{forns2011improving}. Short handles also makes easier the evaluation of the stretching contributions. In fact, the large stiffness of short handles as compared to the trap stiffness, $k_{\text{h}}\gg k_{\text{b}}$, implies that $k_{\text{eff}}\simeq k_{\text{b}}$ to first order. As the trap stiffness itself can be considered nearly force independent $k_{\mathrm{eff}}^{\text{F}}$ is, therefore, constant along the folded branch, and the effective stiffness approximation (\ref{eq:wbhandles2}) becomes applicable.

We tested this approach using a 20 bp DNA hairpin ending in a tetraloop (sequence in top panel of figure \ref{fig:CD4SH}(a)) flanked by two dsDNA handles, each one of 29 bp. The assembled molecular construct (DNA hairpin + handles) shown in figures \ref{fig:CartoonExt}(a),(b), is repeatedly pulled between $\lambda_{0}$ and $\lambda_{1}$. In the forward (reverse) process the system starts in thermal equilibrium at $\lambda_{0}$ ($\lambda_{1}$) and it is driven out of equilibrium following a predetermined protocol $\lambda_{\text{F}}(t)$ ($\lambda_{\text{R}}(t)$) until $\lambda_{1}$ ($\lambda_{0}$) is reached. For each experimental realization the work $W$ is calculated according to \eqref{eq:work}. Note that, in the force range at which the molecule typically unfolds and refolds (12 - 17 pN in figure \ref{fig:CD4SH}(a)), the FDCs are linear in force (inset of figure \ref{fig:CD4SH}(a)). Therefore, the conditions required to use the effective stiffness method are fulfilled \eqref{eq:wbhandles2}.

In figure \ref{fig:CD4SH}(b) we show the F and R work distributions calculated for two pulling speeds (6 and 16 pN/s) in the same integration range. According to the CFT \eqref{eq:CFT}, the work value at which both distributions cross (black solid points) equals to $\Delta G$. Note that, since the integration range is the same, $\Delta G$ does not change with pulling speed, as it is required for an equilibrium quantity. We emphasize the validity of the CFT by plotting the function $\log P_{F}(W) / P_{R}(-W)$ as a function of $W$ in $\kt$ units. According to \eqref{eq:CFT}, this function is linear in $W$ with slope 1 and with a y-intercept equal to $\Delta G$ (both in $\kt$ units). As expected, the experimental data (solid points) satisfy the previous relation (see inset of figure \ref{fig:CD4SH}(b), where the solid line is a linear fit to the experimental data).

Once we have measured $\Delta G$ using the CFT, we subtract the stretching contributions to recover $\Delta G_{0}$. According to \eqref{eq:gtot}, we have:

\begin{equation}
    \label{eq:g0}
    \Delta G_{0} = \Delta G - \Delta W_{\text{m}} - \Delta W_{\text{b}} - \Delta W_{\text{h}} \, .
\end{equation}

The term $\Delta W_{\text{m}}$ is calculated using \eqref{eq:wmol}. In order to model the ssDNA elastic response (i.e. $f(x_{\text{ss}})$ in \eqref{eq:wmol}), we use the WLC model \eqref{eq:wlc} with a persistence length $P$ equal to 1.35 nm and an interphosphate distance $d_{\text{b}}$ equal to 0.59 nm/base \cite{alemany2014determination}, so that $L_{c} = (2 n_{\text{bases}} + 4) \times 0.59$ nm/base $\approx 26$ nm. On the other hand, the term $\Delta W_{\text{h}} + \Delta W_{\text{b}}$ is calculated using the effective stiffness method \eqref{eq:wbhandles2} with $k_{\text{eff}}^{\text{F}} = 0.065 \pm 0.002$ pN/nm (obtained by a linear fit of the FDCs, see inset in \ref{fig:CD4SH}(a)). 

In table \ref{table:cd4sh&cd4lh} we report the values we obtained for $\Delta G$, $\Delta G_{0}$, as well as the aforementioned stretching contributions.

\begin{table}[h!]
\centering
\begin{tabular}{c|cccc}
& $\Delta G$ [$\kt$] & $\Delta W_{\text{m}}$ [$\kt$] & $\Delta W_{\text{h}} + \Delta W_{\text{b}}$ [$\kt$] &  $\Delta G_{0}$ [$\kt$] \\ \hline
DNA short & 295 $\pm$ 1 & 17.4 $\pm$ 0.5 & 225 $\pm$ 4 & $52 \pm 4$\\
RNA long& 342 $\pm$ 2 & 28 $\pm$ 1 & 244 $\pm$ 5 & $70 \pm 5$
\end{tabular}
\caption{\textbf{Fluctuation theorem and stretching contributions for CD4 DNA hairpin (short handles) and CD4 RNA hairpin (long handles).} \textbf{(DNA short, first row)} Reported energies for the integration range [$\lambda_{0}, \lambda_{1}$]=[20, 80] nm corresponding to a force range $(f_{\text{min}}, f_{\text{max}}) = (13, 17)$ pN. \textbf{(RNA long, second row)} Reported energies for the integration range [$\lambda_{0}, \lambda_{1}$]=[30, 85] nm corresponding to a force range $(f_{\text{min}}, f_{\text{max}}) = (18, 22)$ pN. Error bars obtained after averaging the results over four (DNA short) and five (RNA long) molecules at two pulling speeds, respectively.}
\label{table:cd4sh&cd4lh}
\end{table}

Results for $\Delta G_{0}$ are in very good agreement with the theoretical ones obtained using the nearest-neighbour model for DNA either using Mfold parameters ($\Delta G_{0} = 51 \kt$) \cite{zuker2003mfold} or the ones derived from unzipping experiments ($\Delta G_{0} = 48 \kt$) \cite{huguet2010single}. 
\subsection{Long handles: the CD4 RNA hairpin.}\label{sec:long_handles-CD4RNA}


In what follows, we first discuss the characteristics of long handles in subsection \ref{sec:long-handles-intro}, explaining why sometimes the effective stiffness method can be applied, while other times it cannot. To illustrate the two distinct situations, we first present in section \ref{sec:cd4rna} a scenario based on the CD4 RNA hairpin, where the effective stiffness method is applicable with long handles, just as with short handles (Sec. \ref{sec:cd4dna}). Secondly, the development of a general approach for long handles, beyond the effective stiffness approximation, is covered in section \ref{sec:beyond-effective-stiffness} and exemplified with the CD4L12 RNA hairpin.


\subsubsection{Characteristics of long handles.}\label{sec:long-handles-intro}

Long handles, $\sim$ 500 bp each, typically represent a bigger challenge than their short counterpart because they are significantly softer. This implies that long handles stiffness features a  noticeable force dependence $k_{\text{h}} = k_{\text{h}}(f)$, especially in the lower range of forces experimentally accessible with LOT. Moreover, the magnitude of $k_{\text{h}}$ is now lower and typically comparable to the trap stiffness, $k_{\text{h}} \sim k_{\text{b}}$. Thus, since $k_{\text{eff}}^{\text{F}} \simeq (k^{-1}_{\text{h}} + k^{-1}_{\text{b}})^{-1}$, the term $k_{\text{h}}$ significantly contributes to $k_{\text{eff}}^{\text{F}}$. This, together with the clear force dependence of $k_{\text{h}}$, implies in turn that the effective stiffness is not constant but depends on force: $k_{\text{eff}}^{\text{F}} = k_{\text{eff}}^{\text{F}}(f)$. Consequently, upon calculating stretching contributions, the terms $\Delta W_{\text{b}}$, $\Delta W_{\text{h}}$ need to be evaluated more carefully. At closer inspection, however, the use of long handles does not invalidate \textit{per se} the effective stiffness approximation \eqref{eq:wbhandles2}. The validity of \eqref{eq:wbhandles2} relies on the assumption that $k_{\text{eff}}^{\text{F}}$ is constant over the integration range $[\lambda_{0}, \lambda_{1}]$. Indeed, in many situations, such as with CD4 RNA hairpin, the actual integration range occurs at forces high enough so that $k_{\text{h}}\gg k_{\text b}$ and $k_{\text{eff}}^{\text{F}}$ can be taken as constant. Whenever this assumption does not hold another approach must be used. There are two typical scenarios. On the one hand, if the integration range is large (e.g. for molecules featuring a pronounced hysteresis), the force-dependence of the stiffness $k_{\text{eff}}^{\mathrm{F}}=k_{\text{eff}}^{\mathrm{F}}(f)$ cannot be neglected (note that even if $k_{\mathrm{eff}}^{\mathrm{F}}$ changes marginally from pN to pN, the overall change on the whole integration range can be significant). On the other hand, if we reach low enough forces (e.g. by using a molecule that refolds at very low forces), the effective stiffness also exhibits force dependence. Indeed at low forces  $k_{\mathrm{h}} \ll k_{\mathrm{b}}$, hence $k_{\rm eff}^{\mathrm{F}} \sim k_{\mathrm{h}}$, and as $k_{\mathrm{h}}=k_{\mathrm{h}}(f)$ is steep at low $f$, so will $k_{\rm eff}^{\mathrm{F}}$ be.

Provided that the handle stiffness $k_{\text{h}}(f)$ and the force stiffness of the trap $k_{\text{b}}(f)$ are known with a good precision, the integrals in \eqref{eq:wbwh} can in principle be carried out easily, irrelevantly of $k_{\text{eff}}^{\text{F}}$ being non-constant. This corresponds however to an idealized scenario which rarely occurs in practice. To begin with, the elastic properties of the handles are typically characterized for a few ionic and temperature ranges, and often, parametrizations are lacking for generic experimental situations (non-standard buffer, unusual temperatures). Furthermore, for certain types of handles, such as DNA-RNA hybrids, there is a missing gap regarding the elastic description in the literature. But the largest problem comes by far from the usually imprecise knowledge of the optical trap stiffness, since the slightest deviations in the expected value of $k_{\text{b}}$ can have a very significant energetic impact. For instance, a modest deviation of $5\%$ from $k_{\text{b}} = 0.08$ pN/nm to $k_{\text{b}} = 0.075$ pN/nm, results in a change of a dozen $\kt$ in $\Delta W_{\text{b}}$ when integrated between 2 an 12 pN. Changes in the value of $k_{\text{b}}$ and even non-linear force corrections in $k_{\text{b}}(f)$ do inevitably occur in LOT, not only on a day-to-day basis (depending on the laser focusing, alignment, power, intensity or temperature) but also within the same day on a molecule-to-molecule basis, since the beads used for performing experiments can usually slightly vary in size, and  the trap stiffness directly depends on this. A slight force dependence in $k_b(f)$ also occurs if the optical plane of the bead shifts with force. Hence, we see that $k_{\text{h}}(f)$ and $k_{\text{b}}(f)$ are usually not characterized precisely enough for the integrals in \eqref{eq:wbwh} to be computed reliably.

To address the aforementioned issues, we will introduce in  section \ref{sec:beyond-effective-stiffness} a novel methodology to retrieve the optimal stiffness profile $k_{\text{b}}(f)$ and $k_{\text{h}}(f)$ directly from the FDCs obtained in pulling experiments with LOT. 
Before doing so, let us however show an example where long handles and the effective stiffness approximation go in pair: the CD4 RNA hairpin.

\subsubsection{Stretching contributions and folding free energy of CD4 RNA.}\label{sec:cd4rna}

\begin{figure}[h]
\centering
\includegraphics[width=\textwidth]{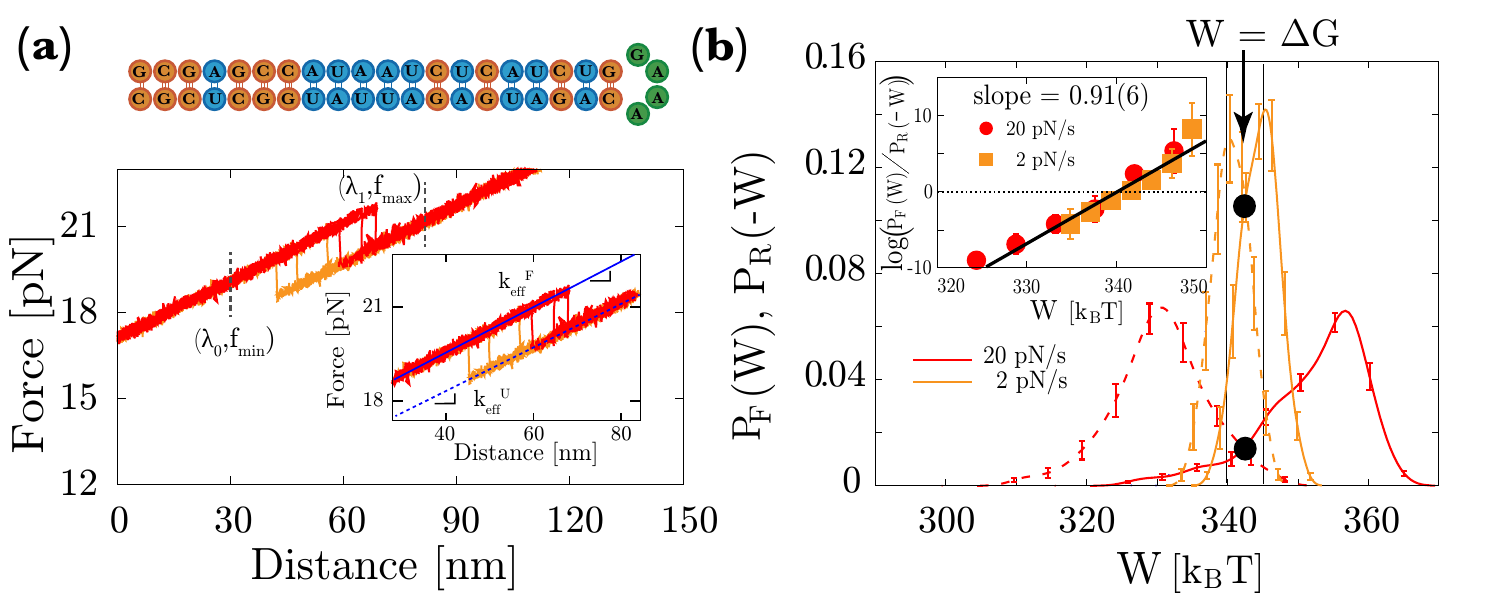}
\caption{\label{fig:CD4LH} \textbf{Free-energy recovery CD4 RNA hairpin with long handles} \textbf{(a).} Sequence of CD4 RNA (top panel). FDCs and integration range for the work $W$ (bottom panel). Visual evidence of the linearity of the FDCs in the integration range (inset) plus linear fits to the folded (solid line in the inset) and the unfolded branches (dashed line in the inset). \textbf{(b).} Forward (solid lines) and reverse (dashed lines) work distributions for two different pulling speeds calculated in the integration range indicated in (a) panel. Crossing points between work distributions are tagged as solid points. The CFT verification is shown as inset. Error bars have been obtained using the Bootstrap method.}
\end{figure}

The effective stiffness method can be applied to the CD4 RNA hairpin which is a molecule showing nearly reversible folding-unfolding kinetics at the accessible pulling speeds \cite{collin2005verification,bizarro2012non}. The molecule has the same sequence as hairpin CD4-DNA presented in Section \ref{sec:cd4dna} but replacing thymines by uracils (top panel of figure \ref{fig:CD4LH}(a)). In the present case, the RNA hairpin is inserted between two $\sim$500 bases-long hybrid RNA/DNA handles \cite{wen2007force}. Thus, the molecular construct is formed by the RNA hairpin plus the two long hybrid handles. Pulling experiments were performed analogously as described in section \ref{sec:cd4dna}. 

Due to the narrowness of the region in the FDCs (figure \ref{fig:CD4LH}(a), bottom panel) where folding-unfolding events of CD4 RNA take place, the effective stiffness $k_{\text{eff}}^{\text{F}}$ remains fairly constant over the force range experimentally probed. This linearity of the FDCs is evidenced  in the inset of figure \ref{fig:CD4LH}(a) and justifies the use of the effective stiffness approximation. By fitting the FDCs slopes in the highlighted region, we obtain a value for $k_{\text{eff}}^{\text{F}}$ equal to $0.067 \pm 0.001$. 

Next, we integrate all FDCs in the range, [$\lambda_{0}, \lambda_{1}$] = [30, 85] nm, which corresponds to the force interval $(f_{\text{min}}, f_{\text{max}}) = (18, 22)$ pN. As we did in section \ref{sec:cd4dna}, the F and R work distributions are calculated for two pulling speeds (2 and 20 pN/s) and are shown in figure \ref{fig:CD4LH}(b). Note that the crossing point between both distributions corresponds to the work value equal to $\Delta G$. The CFT \eqref{eq:CFT} is satisfied for CD4 with long handles, as can be seen in the inset of figure \ref{fig:CD4LH}(b). We can thus subtract from the obtained $\Delta G$ the stretching contributions $\Delta W_{\text{h}} + \Delta W_{\text{b}}$ using the effective stiffness method, along the exact same lines as in \ref{sec:cd4dna}. As a last step, the term $\Delta W_{\text{m}}$ in \eqref{eq:gtot} is calculated using the WLC model \eqref{eq:wlc} with $P = 0.75$ nm and an interphosphate distance $d_{\text{b}}$ equal to 0.665 nm/base, so that $L_{c} \approx 29$ nm, higher than for the CD4 DNA molecule. 

We report in table \ref{table:cd4sh&cd4lh} the values we obtained for $\Delta G$, $\Delta G_{0}$, as well as for the stretching contributions. The measured value for $\Delta G_{0}$ ($70\pm 5 \kt$) is compatible with the previous single-molecule measurements obtained in LOT assays at 100mM Tris HCl pH 8 and 1 M NaCl ($\Delta G_{0} \approx 65$ $\kt$) \cite{bizarro2012non} and with the Mfold prediction ($\Delta G_{0} = 68$ $\kt$) \cite{zuker2003mfold}. We conclude that the effective stiffness approximation is valid for determining folding free energies from irreversible work measurements if the integration range is narrow enough so that FDCs along the folded branch have constant slope in such range (i.e. the effective stiffness $k_{\text{eff}}^{\text{F}}$ can be taken as constant).

\section{Beyond the effective stiffness method}\label{sec:beyond-effective-stiffness}

In the previous sections we introduced the effective stiffness method, testing its reliability in addressing the analysis of both short and long handles. We also gave evidence that its validity is limited to the case of a linear elastic response and that when this condition is not fulfilled a more general methodology becomes necessary. This is the subject covered by section \ref{sec:non-trivial-case-methodology} where we present a novel technique going beyond the effective stiffness approximation.  Then, in section \ref{sec:RNACD4L12} we present an application of this method to the case of CD4L12, a dodecaloop RNA hairpin exhibiting large hysteresis in pulling experiments.


\subsection{Estimating the stretching contributions in the general case.}\label{sec:non-trivial-case-methodology}

As can be seen in (\ref{eq:ext_tot}) the force-extension profile $\lambda(f)$ depends on $x_{\text{b}}$ and $x_{\text{h}}$, and these are, by definition, related to the stiffness through:
\begin{equation} \label{eq:stiffness-ext}
    x_i(f) = \int_{0}^{f} k_{i}^{-1}(f')df' \, , \qquad \frac{dx_i}{df} = \frac{1}{k_{\text{i}}(f)}\qquad \text{for} \, i=b,h \, .
\end{equation}
This hints at the fact that FDCs (i.e. the $\lambda(f)$ profile) might allow us to retrieve the stiffness profiles needed to estimate the elastic energy contributions from bead and handles in \eqref{eq:wbwh}. To realize this in practice, we must assume the elastic response of the trap and the one of the handles can be parametrised by some reasonable physical model. Starting with the handles, we will assume that the extensible WLC model (ext-WLC) is a good description.
\begin{equation} \label{eq:handle-stiff}
    k_{\text{h}}(f) = k_{\text{h}}^{\text{ext-WLC}}(f;\{P,d_{\text{b}},Y\})  \, , \qquad \qquad x_{\text{h}}(f) = x_{\text{h}}^{\text{ext-WLC}}(f;\{P,d_{\text{b}},Y\}) \, ,
\end{equation}
where we introduced the usual WLC elastic parameters (i.e. persistence length $P$, Young modulus, $Y$, and monomer length $d_{\text{b}}$). Then, we can either model the trap stiffness as constant, or as a linear function of force:
\begin{equation} \label{eq:trap-stiff}
    k_{\text{b}}(f) = k_{\text{b},0} + \alpha f \, , \qquad  \qquad x_{\text{b}}(f) = \frac{1}{\alpha} \log \left(1 + \frac{\alpha}{k_{b,0}}f \right) \, ,
\end{equation}
where $\alpha$ quantifies the linear dependence and $k_{\text{b},0}$ is the stiffness at zero force ($x_{\text{b}}(f)$ is obtained by integrating as in \eqref{eq:stiffness-ext}).

Note that we can rewrite \eqref{eq:ext_tot} as:
\begin{equation}
\label{eq:tot-ext-compact}
    \lambda(f) = x_{\text{h}}(f) + x_{\text{b}}(f) + x_{\text{d}}(f)\delta_{N} + x_{\text{ss}}(f)\delta_{U} + \lambda_0 \, ,
\end{equation}
where we used a delta-Kronecker-like notation ($\delta_{N(U)}=1$ if the molecule is in the Native (Unfolded) state and zero otherwise) and explicitly introduced the offset $\lambda_0$, which accounts for the fact that the molecular extension is always measured with respect to the micropipette. If we now rewrite the explicit dependence with respect to our model parameters, \eqref{eq:tot-ext-compact} becomes:
\begin{equation} 
\begin{split} \label{eq:fit-ext}
\lambda(f) &= x_{\text{h}}(f;\{P,d_{\text{b}},Y\}) \, + \, x_{\text{b}}(f;\{\alpha,k_{\text{b},0}\}) \, +  \, x_{\text{d}}(f)\delta_{N} + x_{\text{ss}}(f)\delta_{U} \, + \, \lambda_0 \\
        &\equiv  \mathcal{M}(f;\{P,d_{\text{b}},Y,\alpha,k_{\text{b},0},\lambda_0\}) \, ,
\end{split}
\end{equation}

where we have denoted $\mathcal{M}$ as the overall model underpinning the $\lambda(f)$ response. As \eqref{eq:ext_tot} illustrates, the knowledge of a handful of physical parameters fully determines the FDC for the N and U branches. The key idea behind our methodology is that the inverse implication is also true: knowing $\lambda(f)$ and given $\mathcal{M}$ we can extract $P,d_{\text{b}},Y,\alpha,k_{\text{b},0}$ through \eqref{eq:fit-ext}, using a least squares method, a Bayesian approach, or generally any regression method. Once these parameters are fitted, we can recompute $k_{\text{h}}(f)$ and $k_{\text{b}}(f)$ for all $f$ using the models in \eqref{eq:handle-stiff} and \eqref{eq:trap-stiff} eventually obtaining the stretching contributions through numerical integration of \eqref{eq:wbwh}. Crucially, this can be done without any \textit{a priori} knowledge of the parameters of the experimental setup.

In practice, however, the fitting procedure requires a FDC featuring enough curvature to be able to constrain the model, and even so, the number of parameters to fit is too large for a 2-dimensional curve, so that some additional considerations must be taken into account. Firstly, reasonable bounds/priors on the allowed values for the parameters must be set. Secondly, it is convenient to assume that certain parameters play a minor role in the overall FDC shape (such as $Y$) or are characterized well enough (e.g. the monomer length for dsDNA) to be fixed at some nominal value and not fitted. Thirdly, computing the handles extension $x_{\text{h}} = x_{\text{h}}(f)$ using the extensible WLC can be slow and numerically inaccurate as it normally requires to perform a numerical inversion of $f = f(x_{\text{h}})$. To address this, we introduce in \ref{app:inversionWLC} a formula to explicitly invert the WLC which can then be used in \eqref{eq:fit-ext}. Fourthly, to get as many points as possible to constrain the fit, we have aligned all the FDCs in the starting point so they share an identical $\lambda_0$ offset (i.e. 'const' in \eqref{eq:ext_tot}(a,b)). After all these steps, fitting $\lambda(f) = \mathcal{M}(f;\{P,d_{\text{b}},Y,\alpha,k_{b,0},\lambda_0\})$ is affordable.

In the following section we will show a concrete examples of the FDCs fitting procedure and its application to extract the stretching contributions.

\subsection{Application to the specific example of CD4L12 RNA hairpin.}\label{sec:RNACD4L12}
\begin{figure}[h!]
\includegraphics[width=\textwidth]{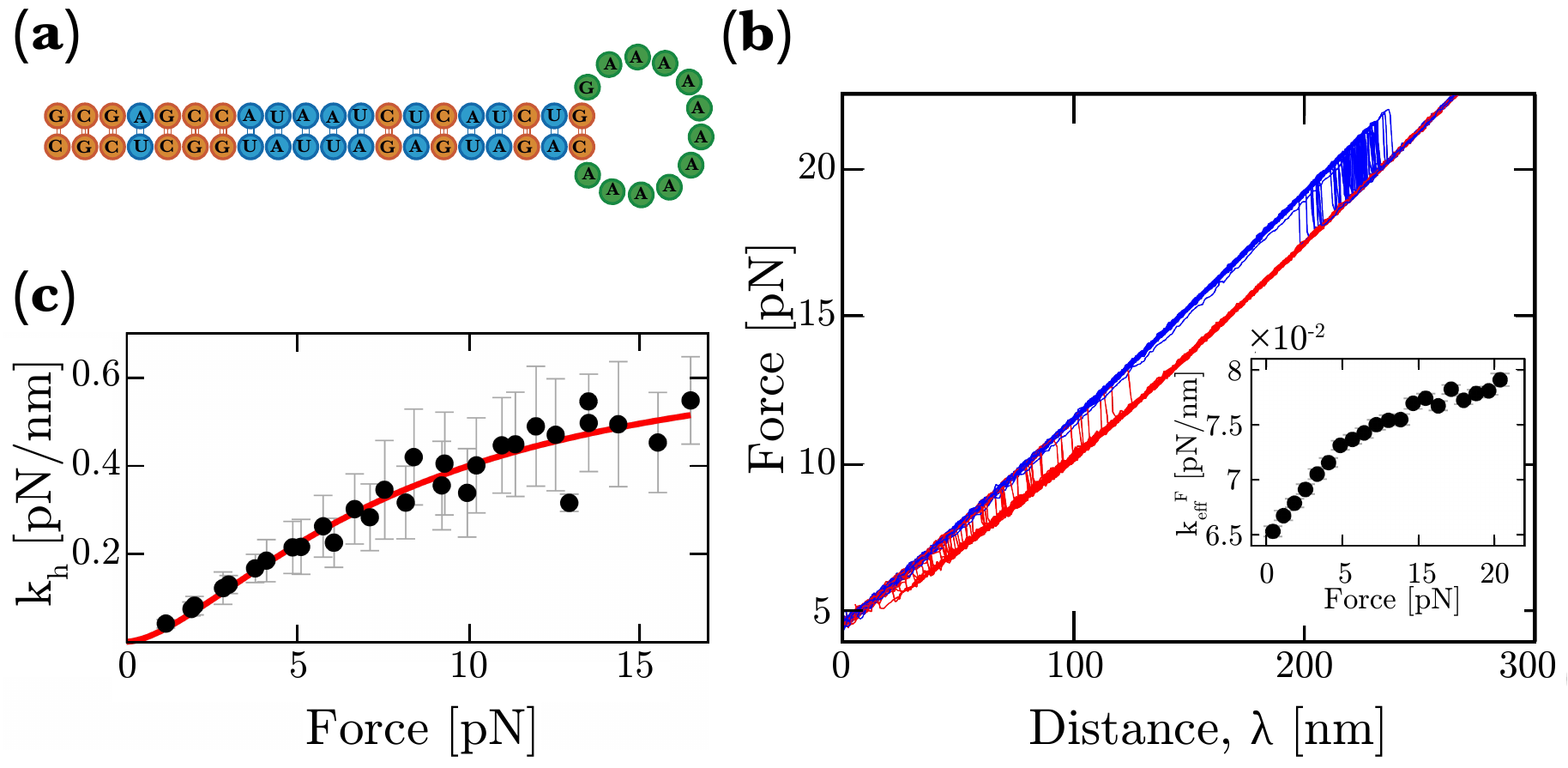}
\caption{\label{fig:non-trivial-case}\textbf{Free-energy recovery CD4L12 RNA hairpin.} \textbf{(a).} Sequence and secondary structure of CD4L12 RNA \textbf{(b).}  Aligned FDCs folding (red) and unfolding (blue) for a given molecule pulled at 100 nm/s. \textit{Inset}: Effective stiffness profile measured along the folded branch. \textbf{(c).} Stiffness profile of the hybrid DNA-RNA handles which form the molecular construct used with CD4L12 and CD4 RNA \cite{bizarro2012non}. Data points have been obtained using the high frequency power-spectrum method described in \cite{forns2011improving}. The red line is a fit of the extensible WLC model, yielding $P=20\pm4$nm and $Y=200\pm14$ pN ($d_{\text{b}}$ was not fitted but fixed to the interphosphate distance of A-form RNA, $d_{\text{b}} = 0.27$ nm \cite{wen2007force}). }
\end{figure}

The effective stiffness method may work well when the range of force integration is not too large. This condition is met in molecules exhibiting mild hysteresis. For molecules showing large hysteresis in pulling cycles the limits of integration $f_{\text{min}}$ and $f_{\text{max}}$ are far away and the effective stiffness $k_{\text{eff}}^{\mathrm{F}}$ cannot be considered constant anymore. Here we present results for an RNA molecule (CD4L12) falling in this category and present a general procedure to extract the free energy of formation. CD4L12 shares the same stem than the previously discussed CD4 RNA in section (\ref{sec:cd4rna}), but with the original tetraloop replaced by a dodecaloop (i.e. 12-loop bases), see sequence in figure \ref{fig:non-trivial-case}(a). A large loop yields a larger entropic barrier for refolding and large hysteresis in the FDC. Pulling experiments were performed as described in section (\ref{sec:cd4dna}), with a pulling speed of 100 nm/s and 300 nm/s and a buffer containing $4$ mM MgCl, $50$ mM NaCl, and $10$ mM Tris. The values of $P=0.75 $ nm and $d_{\text{b}} = 0.665 $ nm were used to describe the elastic properties of the ssRNA for this buffer \cite{bizarro2012non}.

As can be seen in figure \ref{fig:non-trivial-case}(b), CD4L12  behaves as a two-state system being either folded or unfolded along the FDCs. As expected, pulling cycles feature large hysteresis, with a maximal difference of nearly $20$ pN between the lowest folding and largest unfolding force rips. In order to compute the work needed for the CFT \eqref{eq:CFT}, we must integrate the area under the FDC within a large force range with a very low $f_{\text{min}}$. It is clear that in this case the constant stiffness approximation described in section \ref{sec:long-handles-intro} does not apply, as shown in the inset of figure \ref{fig:non-trivial-case}(b) where $k_{\text{eff}}^{\text{F}}$ markedly changes with force. To estimate the stretching contributions we follow the previous subsection \ref{sec:non-trivial-case-methodology} and \eqref{eq:fit-ext} to obtain $\Delta W_{\text{b}}$, $\Delta W_{\text{h}}$, and, from \eqref{eq:g0}, the value of $\Delta G_0$.

\begin{figure}[h!]
\includegraphics[width=\textwidth]{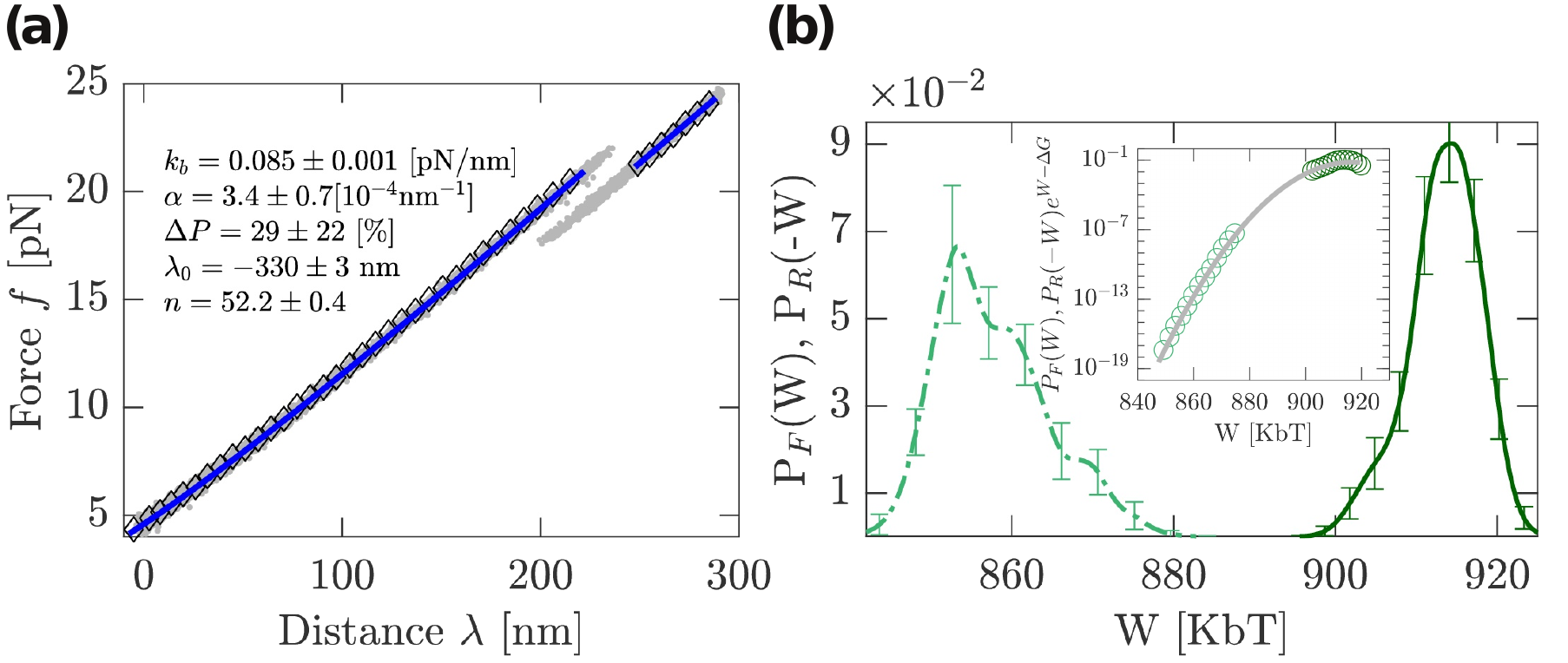}
\caption{\label{fig:CD4L12-model-fit}\textbf{Fitting the folded and unfolded branches of CD4L12 RNA hairpin.} \textbf{(a).} Solid blue line is an example of curve fitting based on \eqref{eq:correc-model-used-in-practice}. Data points used for the fit are the black diamonds. They are obtained by smoothing and filtering the gray dots, which are themselves obtained by aggregating the unfolding FDCs of different pulling cycles from figure \ref{fig:non-trivial-case}b \textbf{(b).} Example of forward and reverse (solid and dashed lines) work distributions for the same molecule pulled at 100 nm/s. Due to the large hysteresis, work distributions do not overlap. \textit{Inset:} Illustration of the \emph{matching} method to retrieve $\Delta G$ by imposing continuity between $P_F(W)$ (light green) and $P_R(-W)e^{(W-\Delta G) / \kt}$ (dark green) in log-normal scale. Solid grey line is the fitted Gaussian, see \cite{alemany_mechanical_2016} for details.}
\end{figure}

In order to carry out the fit prescribed by \eqref{eq:fit-ext}, we need to introduce some further assumptions to simplify the problem. Regarding the hybrid DNA-RNA handles, we use the value of the interphosphate distance $d_{\text{b}} = 0.27$ nm of A-form RNA and Young modulus $Y = 200$ pN obtained by fitting the stiffness of the handles profile (figure \ref{fig:non-trivial-case}(c)).  While changes in $d_{\text{b}}$ only moderately affect the overall curvature of $k_{\text h}$ (but they impact the overall contour length, an effect already captured by fitting $\lambda_0$), changes in $Y$ do not. Hence fixing these two values gives a better constrained model. For the persistence length of the handles $P$ it is convenient to fit the deviation $\Delta P$ (in \%) with respect to a plausible expected nominal value $P_0$, i.e. $P_{\text{eff}} = P_0 \left( 1 + \Delta P \right) $, that we take from the fit in figure (\ref{fig:non-trivial-case}) as $P_0=20$ nm. Lastly, we also include the number of nucleotides $n$ released in the transition between the folded and the unfolded branches as an extra free parameter of the model. We are thus eventually left with 5 free parameters which we fit (\ref{eq:tot-ext-compact},\ref{eq:fit-ext}) using a standard non-linear least square regression (Levenberg-Marquardt): 
\begin{equation}\label{eq:correc-model-used-in-practice}
    \lambda(f) = \mathcal{M}(f) = \mathcal{M}(f; \{k_{\text{b,0}}, \alpha, \Delta P, \lambda_0,n)\})~~~. \,
\end{equation}

An example of such fitting procedure is shown on figure \ref{fig:CD4L12-model-fit}(a). As can be seen, the agreement between the experimental points and the reconstructed curve is remarkable. Furthermore, all the values obtained from the fit dovetail with prior expectations. Firstly, the value of $n$ matches with the expected number of released nucleotides (i.e. 52). Secondly, the zero-force trap stiffness $k_{\text{b}}$ falls in expected range \cite{forns2011improving}. Thirdly, the force-dependence parameter $\alpha$ of the trap stiffness is of the same order of magnitude than values already reported in the literature for similar LOT settings \cite{forns2011improving}. Fourthly, $\Delta P$ is small so $P$ is reasonably close to the assigned nominal value $P_0$. Another good generic indicator is the very low error on the fitted parameters, hinting at a well-constrained model; a fact that is further confirmed by the observation that in the correlation matrix of the fit, most off-diagonal entries are near-zero (details not shown). We must finally stress that the choice of free parameters in (\ref{eq:correc-model-used-in-practice}) is convenient for the considered situation but is by no means customary. In a context where the trap would be well characterized and the handles would not, we may have for instance fixed $k_{\text{b}}$ but fitted $d_{\text{b}}$. Equation \eqref{eq:fit-ext} can be adapted at will, depending on the requirement.

With the fitted values of $\alpha$, $k_{\text{b,0}}$ and $\Delta P$ in hand and our assumptions for $Y$ and $d_{\text{b}}$ (legitimated retrospectively by the agreement of the fit in figure \ref{fig:CD4L12-model-fit}), we are now in a position to precisely establish the profiles of $k_{\text{h}}(f)$ an $k_{\text{b}}(f)$ through the use of equations \eqref{eq:stiffness-ext}, \eqref{eq:handle-stiff} and \eqref{eq:trap-stiff}. We can now quantify the terms $\Delta W_{\text{b}}$ and $\Delta W_{\text{h}}$ using \eqref{eq:wbwh} and $\Delta G$ using the FT. These numbers together with \eqref{eq:g0} allow us to extract $\Delta G_0$.

Figure \ref{fig:CD4L12-model-fit}(b) shows the work distributions $P_F(W)$ and $P_R(-W)$ obtained from the FDC (figure \ref{fig:non-trivial-case}(b)). The very pronounced hysteresis and the large value of the average dissipated work in a pulling cycle (about 60 $\kt$) is such that F and R work distributions lie far apart without overlapping. Previous methods based on the overlapping of F and R work distributions are not applicable and an alternative approach must be used, such as the Bennett acceptance ratio \cite{bennett1976efficient} and the ``matching method''. This last method consists in finding the optimal $\Delta G$ value so that $P_F(W)$ is the analytical continuation of $P_R(-W)e^{(W-\Delta G)/\kt}$. This procedure is graphically illustrated in the inset of figure \ref{fig:CD4L12-model-fit}(b) and further explained in \cite{alemany_mechanical_2016}. Results obtained for different molecules are shown in table \ref{table:dg-cd4l12}. We note that the values of $\Delta G$ obtained with the two methods yield compatible results (\emph{matching} being systematically 3-5 $\kt$  lower than \emph{Bennett}). Our estimated value $\Delta G_0 = 67 \pm 2\kt$ is not far from the Mfold prediction ($\Delta G_0 = 63$ $\kt$) showing the reliability of the approach. 

\begin{table}[h!]
\centering
\begin{tabular}{cccccc}
$\Delta G_{\text{Bennet}}$ [$\kt$] & $\Delta G_{\text{\emph{Matching}}}$ [$\kt$] & $\Delta W_{\text{m}}$ [$\kt$] & $\Delta W_{\text{b}} + \Delta W_{\text{h}}$ [$\kt$] & $\Delta G_0$ [$\kt$] \\ \hline
1045 $\pm$ 3 & 1040 & 35 & 944 $\pm$ 3 & 66 \\
950 $\pm$ 2 & 947 & 35 & 846 $\pm$ 1 & 68 \\
863 $\pm$ 3 & 859 & 35 & 758 $\pm$ 2 & 70 \\
888 $\pm$ 2 & 886 & 35 & 790 $\pm$ 2 & 63 \\
938 $\pm$ 4 & 935 & 35 & 838 $\pm$ 1 & 66 \\ 
1107 $\pm$ 2 & 1105 & 38 & 1003 $\pm$ 2 & 68 &  \\ \hline
  &   &   &  \textbf{Mean:}   & $\mathbf{ 67\pm2 } $ $\mathbf{\kt}$
\end{tabular}
\caption{\textbf{Fluctuation theorem and stretching contributions for CD4L12 RNA hairpin with long handles.} Overview of the values of $\Delta G$, the stretching corrections, and the final $\Delta G_0$ estimate for 6 different molecules. All values are given in $\kt$. $\Delta G_{\text{Bennet}}$ and $\Delta G_{\text{\emph{Matching}}}$ provide two ways to extract $\Delta G$ using the CFT. The value of $\Delta G_0$ is obtained through \eqref{eq:gtot} using the value of the Bennett estimate. The last line corresponds the only experimental setting in which the pulling speed is 300 nm/s, all the other results were obtained at 100nm/s.}
\label{table:dg-cd4l12}
\end{table}

We want to stress the sensitivity of the value of $\Delta G_0$ on the accurate estimation of the stretching contributions which, being one order of magnitude larger, can lead to inconsistent results. Had we used a methodology assuming 'average' or 'standard' stretching contributions, we would have obtained erroneous numbers. Consider for instance subtracting the average value  $\left\langle \Delta W_{\text{b}} + \Delta W_{\text{h}} + \Delta W_{\text{m}} \right\rangle=898\kt$ derived from table \ref{table:dg-cd4l12} to the highest and  the lowest estimates of $\Delta G$ shown in the same table: it results in two widely off values $\Delta G_0 = 1107 -898 = 209 \kt$ and $\Delta G_0 = 863 - 898 = - 35 \kt$. Therefore a tailored molecule-to-molecule estimation of the stretching contribution is absolutely essential for molecules like CD4L12 where the effective stiffness approximation cannot be used.

\section{Conclusions}\label{sec:conclusions}

We have presented a brief tutorial on the approaches commonly used to extract folding free energies of single molecules pulled with optical tweezers in unzipping assays. A recurrent issue in these calculations is the large magnitude of the stretching contributions to the full free-energy difference measured in a pulling experiment using the CFT. Such contributions arise from the experimental setup and include the optical trap, the elastic stretching of the handles used in the molecular construct and the extension release of the unfolded polymer. A great simplification in the analysis of these correction terms can be be performed when the effective stiffness of the experimental system can be approximated as constant, as we saw in section \ref{sec:effectivestiffness-examples}. In this so-called effective stiffness approximation a single parameter $k_{\text{eff}}^{\text{F}}$ suffices to quantify the stretching contributions of handles and trap. We exemplified this case in the study of a DNA hairpin in section \ref{sec:cd4dna}. For long handles the stiffness of the handles turns out to be comparable to that of the trap and a force dependent $k_{\text{eff}}^{\text{F}}$ is apparent. In this case, as we showed in section \ref{sec:cd4rna}, one can still use the effective stiffness approximation if the range of integration to evaluate the work is narrow enough. This is possible if the pulling curves are not too irreversible and forward and reverse work distributions overlap. In contrast, for strong irreversible pulling experiments one needs to accurately characterize all elastic contributions from the experimental setup. Here we have introduced a novel method (section 5) based on least-squares fitting of the elastic response of the folded and unfolded branches. It relies on adapting the elastic parameters extracted from the literature (inter-monomer distance, persistence length, Young modulus) to the experimental data as well as accurately retrieving the stiffness of the optical trap using the same data.

One problem that remains open is the magnitude of the statistical error committed in the estimation of $\Delta G_0$. In fact, $\Delta G_0$ is the difference of two large numbers ($\Delta G$ and the stretching contributions) each with a large error and extracted from the same experimental FDC data. How to combine the errors from these two large quantities remains largely unclear as they are not really uncorrelated. A rule of thumb in single-molecule experiments is that the largest errors come from molecule to molecule experimental variability. It is then recommended to first extract $\Delta G_0$ values for different molecules by subtracting elastic contributions from $\Delta G$ on a single-molecule basis, and then derive the mean value of $\Delta G_0$ and the corresponding statistical error. 

The large contribution of the stretching term \eqref{eq:g0} to the full free energy $\Delta G$ makes the prediction of the (comparably small) value of $\Delta G_0$ a difficult task. This situation is reminiscent of the enthalpy-entropy compensation problem in biochemistry \cite{petruska1995enthalpy,sharp2001entropy}. In this case free-energy differences of intra an intermolecular weak interactions (e.g. folding, binding, allostery, enzymatic reactions and so on) are typically one order of magnitude smaller than entropies and enthalpies, i.e. $\Delta G=\Delta H-T\Delta S$ with $\Delta G\ll \Delta H,T\Delta S$. In this regard, enthalpy-entropy compensation in biochemistry appears to be similar to the $\Delta G$-stretching compensation in force spectroscopy. The analogy is not pure coincidence as the stretching contributions are essentially also of entropic nature and much larger than the bare free-energy difference $\Delta G_0$.  

The methodology we have described should be generally useful and applicable to force spectroscopy studies of single-molecule constructs whenever elastic contributions are present. Applications go beyond the case of measuring folding free energies such as extracting molecular free-energy landscapes \cite{woodside2006direct}  measure ligand binding energies  \cite{camunas2017experimental}, protein-protein and RNA-protein interactions and characterizing heterogeneous molecular ensembles \cite{martinez2019free}. 

{\bf Acknowledgements.} We acknowledge financial support from Grants Proseqo (FP7 EU program) FIS2016-80458-P (Spanish Research Council) and Icrea Academia prizes 2013 and 2018 (Catalan Government). 

\section*{References}

\bibliographystyle{iopart-num}
\bibliography{bibliography}

\newpage

\appendix
\section{WLC Explicit inversion}\label{app:inversionWLC}
The inextensible WLC model described in (\ref{eq:wlc}) gives a very direct way to compute $f = f(x)$, but it is not straightforward to use it to retrieve $x=x(f)$. Although numerical inversion using Mathematica and other software is possible (e.g. as in \cite{broekmans2016dna}) it is useful to have explicit inversion formulae. Hence let us now quickly show that (\ref{eq:wlc}) can be easily inverted to express $z:=x/L_c$ as a function of $f$. We first define the normalized quantity $\tilde{f}=(4P/k_{\text{b}} T)f$. We can then re-write \eqref{eq:wlc} as $ \tilde{f}=\left(1-z\right)^{-2} - 1 + 4z$. By multiplying both sides of the previous by $(1-z)^2$ and by moving all terms to the same side, we obtain:
\begin{equation}
\label{invertedWLC}
0 = z^3 + a_2 z^2 + a_1 z + a_0 \qquad \text{with} \qquad a_2=-\frac{9}{4}-\frac{\tilde{f}}{4}, \quad a_1=\frac{3}{2} + \frac{\tilde{f}}{2}, \quad a_0 =-\frac{\tilde{f}}{4}
\end{equation}
Thus we directly see that obtaining $z$ as a function of $f$ simply  maps to finding the roots of a cubic polynomial -- a problem solved since the 15th century. The approach taken here is the canonical one \cite{abramowitz1965handbook,borwein1995polynomials}. We start defining the following intermediate quantities:
\begin{equation}
R := \frac{9a_1 a_2 - 27 a_0 - 2 a_{2}^{3}}{54} \qquad Q:=\frac{3a_1 - a_{2}^{2}}{9}
\end{equation}
from which we obtain the standard determinant $D$ for cubic equations:
\begin{equation}
D:= Q^3 + R^2
\end{equation}
If $D>0$, there is only one real solution to \eqref{invertedWLC}, and we have to define the following intermediate quantities to express the answer:
\begin{equation}
T:=\sqrt[3]{R+\sqrt{D}} \qquad S:=\sqrt[3]{R-\sqrt{D}}
\end{equation}

(since $D>0$, we also have that $\sqrt{D}$ is real, and thus there is indeed at least one real cubic root for $T$ and $S$). The desired inverse value $z^{*}=z(f)$ is then finally obtained as:
\begin{equation}
z^{*} = -\frac{1}{3}a_2 + S+T
\end{equation}

If $D<0$, there are three real roots to the cubic equation. These roots can be obtained by re-using the quantities $S$ and $T$ defined above, but doing so requires using complex number algebra -- which may not be handy. Instead, we also can define the following intermediate quantity:
\begin{equation}\label{theta-cubic}
\theta := \arccos \left( \frac{R}{\sqrt{-Q^3}}\right)
\end{equation}

From which the three real roots $z_1,z_2, z_3$ can be obtained directly as:
\begin{equation}
z_i = 2\sqrt{-Q} \cos\left(\frac{\theta + \theta_i}{3} \right) - \frac{1}{3} a_2 \qquad \text{with} \qquad \theta_1 = 0, \quad \theta_2 = 2\pi, \quad \theta_3 = 4\pi
\end{equation}

The root of interest is the one lying in the interval $[0,1]$, since we defined above $z=x/L_c$ and a property of the inextensible WLC is that the extension $x$ is always smaller than the contour length $L_c$. Using trigonometric standard formula and the fact that $2\sqrt{-Q}>0$ it is quite easy to verify that $z_1-z_2>0$ and $z_3-z_2\geq0$ for the given range of $\theta$ (which must belong to $[0,\pi]$ by definition of the arccosine), which implies that $z_2$ is the smallest of all the roots. Moreover, we note that all the roots must be positive, since we see in \eqref{eq:wlc} that  $\forall z<0$, $f(z)<0$ and is strictly monotonically decreasing. As all the roots are positive and $z_2$ is the smallest of them, it therefore has to be the one we are looking for, in $[0,1]$, and hence $z_2 = z^{*}=z(f)$ when $D<0$. The previous result also covers the $D=0$ situation, because we then have from \eqref{theta-cubic}, $\theta = 0$, and so we are in the limiting case $z_3=z_2$.

Let us finally note that in the case of the \textit{extensible} WLC, the key difference with the inextensible case is the replacement $L_c \rightarrow L_c(1+f/Y)$ with $Y$ the Young Modulus, i.e. the contour length is now force dependent. It can be shown that this implies the following relationship between the two models:
\begin{equation}
x_{WLC}^{ext} (f) = x_{WLC}^{inext} (f) \left( 1 + f/Y \right)
\end{equation}
and so we see that knowing the explicit inversion for the inextensible model directly yields an explicit formula for the extensible model too.

\end{document}